\documentstyle[12pt,epsf]{article}
\def\fun#1#2{\lower3.6pt\vbox{\baselineskip0pt\lineskip.9pt
        \ialign{$\mathsurround=0pt#1\hfill##\hfil$\crcr#2\crcr\sim\crcr}}}
\def\abstract{\if@twocolumn
\section*{Abstract}
\else \large 
\begin{center}
{\bf Abstract\vspace{-.5em}\vspace{0pt}} 
\end{center}
\quotation 
\fi}
\def\endabstract{\if@twocolumn\else\endquotation\fi}
\def\kok{ {\rm K}^ 0 - \overline{\rm K}{}^0 }
\def\Uscr{\cal U}
\def\Xscr{{\cal X}}

\def\Wscr{\cal W}

\def\Sscr{\cal S}

 \newcount\timehour          
 \newcount\timeminute
 \newcount\timehourminute
\def\daytime{
\timehour=\time\divide\timehour by 60
   \timehourminute=\timehour\multiply\timehourminute by -60
   \timeminute=\time\advance\timeminute by \timehourminute 
   \number\timehour:\ifnum\timeminute<10{0}\fi\number\timeminute}
\def\today{\number\day\space\ifcase\month\or Jan\or Feb\or Mar
   \or Apr\or May\or Jun\or Jul\or Aug\or Sep\or Oct\or 
   Nov\or Dec\fi\space\number\year}

\newcommand{\pslash}
           {\mbox{$ p \hspace{-1ex} \mbox{/} \hspace{-0.08ex} $}}

\begin{document}

\title{\vskip-2.5truecm{\hfill \baselineskip 14pt {{
\small  \\       \hfill \large IFUM--FT 640/99  
}}
\vskip 4truecm}
 {\LARGE {\bf The exact parametrization}}
\vskip 0.5truecm
 {\LARGE {\bf of the neutrino mixing}}}
\vspace{3.5cm}

\author{D. Cocolicchio$^{1,2}$ and M. Viggiano$^{1}$
 \\  \  \\
 $^{1}${\it Dipartimento di Matematica, Univ. Basilicata, Potenza, Italy}\\ 
{\it Via N. Sauro 85, 85100 Potenza, Italy} \\ 
 $^{2}${\it Istituto Nazionale di Fisica Nucleare, Sezione di Milano, Italy}\\
{\it Via G. Celoria 16, 20133 Milano, Italy}
}

\date{}
\maketitle
\vfill

\begin{abstract}
\baselineskip 20pt
\noindent
{\large
We discuss the propagation and the flavour mixing of neutrinos in the context 
of the Quantum Field Theory. We propose an exact parametrization of the neutrino
mixing matrix and its transformation in a simpler form which accounts for a
small $\nu_e$--$\nu_\mu$ and a large $\nu_\mu$--$\nu_\tau$ mixing.
Finally, we comment the differences between the charged quark and neutral lepton
sector and consider the physical situations that require this approach.
}
\end{abstract}
\vfill
\thispagestyle{empty}

\newpage
\pagestyle{plain}
\setcounter{page}{1}
\def\beq{\begin{equation}}
\def\eeq{\end{equation}}
\def\beqa{\begin{eqnarray}}
\def\eeqa{\end{eqnarray}}
\baselineskip 20pt

\leftline{\bf I. The propagation of the neutrino system.}
\parskip 8pt
\noindent
There are several models to justify the propagation of neutrinos with 
flavour mixing. 
As it is well known, if neutrinos $\nu_e$, $\nu_\mu$ and $\nu_\tau$ 
have non-zero masses, they may oscillate into each other
\cite{BP}, analogously to the complex $\kok$ system
\cite{CV}.
The recent experimental results of the
Super--Kamiokande Collaboration \cite{SKam} seem to indicate 
a zenith--angle dependence of the
atmospheric $\nu_\mu$ showers which are depleted as they traverse earth 
as a consequence of $\nu_\mu\leftrightarrow\nu_{\tau}$
transitions and, therefore, in contrast with
the expectations of the Standard Model.
Oscillations have also been invoked to describe the 
appearance of electron neutrinos and antineutrinos in the muon 
(anti)neutrino flux of the LSND 
accelerator experiment \cite{LSND}, to account the $^8$B solar neutrino 
deficit \cite{SSM}, and, at same time, to fulfill the constraints from the 
CHOOZ reactor experiment \cite{CHOOZ}.
Indeed, the observed flux depletions could be ascribed both to
vacuum neutrino oscillations \cite{BP}\ or to a matter resonant
flavour conversion \cite{MSW}.
The oscillation phenomena occur because
weak flavour states $\nu_\ell \in \{ \nu_e,\nu_\mu,\nu_{\tau} \} $ are
produced and detected whereas the propagation is diagonal in the entangled 
propagating eigenstates $\nu_\alpha$, $\alpha=1,2,3$. Flavour eigenstates do not 
necessarily coincide with the propagating eigenstates.  
A matrix $V$ transforms the initial ``flavour" states
$|\nu_\ell\rangle$ into the ``propagation'' eigenstates
$|\nu_\alpha\rangle$, i.e.
\beq\label{transf}
|\nu_\alpha\rangle=V_{\alpha\ell}|\nu_\ell\rangle.
\eeq
In this paper,
we prefer the latin indices $\ell , h$ to denote the flavour states
$\nu_e,\nu_\mu,\nu_{\tau}$ and, the greek letters for
the propagating eigenstates $\nu_\alpha$, $\alpha$=1,2,3.
The appearance of $V^{-1}\neq V^\dagger$ implies that the 
propagation eigenstates are not orthonormal \cite{CV}. 
In general, the time evolution of the flavour states is given by
\beq
|\nu_\ell(t)\rangle= {\Uscr}_{\ell h} \vert \nu_h 
(t_0)\rangle \eeq
and similarly for the propagating right--eigenstates
\beq\label{prop}
|\nu_\alpha(t)\rangle= {\Wscr}_{\alpha \beta}
\vert \nu_\beta (t_0)\rangle
=\delta_{\alpha\beta}\mbox{e}^{-i\lambda_\beta t}|\nu_\beta(t_0)\rangle \; ,  
\eeq 
being $\Wscr$ a diagonal functional matrix in absence of
vacuum regeneration  effects \cite{CV}.
The evolution matrices $\Uscr$ 
and $\Wscr$ are related by the following similarity transformation
\beq
{\Uscr} = V^{-1} {\Wscr} V \quad .
\eeq
%
The time--dependence of the probability amplitude ${\cal A}_{\ell h}={\cal
A}(\nu_\ell \to \nu_h)$ for a generic oscillation of a flavour state
$\vert \nu_\ell \rangle$ into another $\vert \nu_h \rangle$
can be expressed in terms of a linear superposition of the propagating
eigenstates $\vert
\nu_{\alpha} \rangle$
\beq\label{amplitude}
{\cal A}_{\ell h} = {\cal A}(\vert \nu_\ell \rangle \rightarrow \vert 
\nu_h \rangle ) = \sum_{\alpha , \beta} \langle \nu_h \vert \nu_{\alpha}
\rangle    \langle \nu_{\alpha} \vert e^{-iHt} \vert \nu_{\beta} \rangle \langle
\nu_{\beta} \vert \nu_\ell \rangle \; .
\eeq
In relativistic field theory, the propagation amplitude 
$\Delta^{-1}_{\alpha \beta}=\langle \nu_{\alpha} \vert e^{-iHt} \vert \nu_{\beta}
\rangle$ is  usually identified with a matrix Feynman propagator.
In the case of the non-relativistic reduction,
flavour oscillations are intimately connected with the correlations
emerging in the propagation.
In this context, the time evolution operator ${\Uscr} (t, t_0)$ for 
multilevel systems described by an interacting Hamiltonian $H=H_0 + H_I$, can
be  represented by means of the infinite Feynman--Dyson series
\beq
{\Uscr} (t,t_0) = \sum_{n=0}^\infty {\Uscr}_n(t,t_0) \qquad {\rm with} 
\eeq 
\beq
{\Uscr}_n(t,t_0)= {1 \over n!} \left({-i \over \hbar}\right)^n \int_{t_0}^t dt_1 \; 
\dots \int_{t_0}^t dt_n \; T \Big( H_I(t_1) \dots H_I(t_n) \Big) 
\eeq
where we can perform all integrations over the same time interval because of 
the properties of the chronological $T$ ordering operator. Formally, summing
the infinite series, we obtain an exponential form
\beq
{\Uscr}(t,t_0) =  e^{\Omega(t,t_0)} \,.
\eeq
There are a number of advantages in an exponential form of the time
evolution operator. In fact, an operator which is expressed as an exponential
power series of Hermitian operators has the added advantage that formal
unitarity is preserved even if only a finite number of terms is considered.
Of course, the existence of a non Hermitian term in the Hamiltonian 
will also lead to a violation of unitarity, but all that does not 
create additional troubles, with few minor changes only.
The solution $\Omega(t,t_0)$ has been derived in a large variety of 
both analytical and numerical methods.~It was expressed
with mathematical elegance as
a series of successively higher orders of commutators by Magnus
\cite{Magnus}. The Magnus operator $\Omega(t,t_0)$ is just the continuous analog
of the Baker-- Campbell--Hausdorff  formula for discrete operators. This work
was extended by Wilcox \cite{wilcox} and many others \cite{BCH}, who calculated
the Magnus expansion to the fourth order.  Of course, this technique is non
trivial because it gives rise to a (nonlinear)
system of  matrix operator differential equations. However, the problem can be
rooted into the representation theory and in terms of group invariants.
This group theoretical approach has the advantage of being applicable 
to any physical system and it provides, for example, the matrix 
representation of $\Omega$ by means of a faithful representation of 
the generators of the relevant Lie group of the model \cite{BCH}.
However, in order to consider higher orders than the fourth,
a convenient representation could be attained only in terms of Cauchy
integrals \cite{birula} with a generalization of the Sylvester expansion
formula. Indeed, this Schr\"odinger--like wave--function formalism has been
already applied to derive the time evolution for three generations of neutrinos
\cite{baldini}, and found that the essence of the problem
depends on the mixing and the hierarchy of the spectrum of the neutrino
system. Of course, the presence of 
neutrino correlations \cite{cohe}\  can alter the validity of 
this approximation, 
and probably, only a density matrix formalism becomes reliable
\cite{dens}.
It is worth noting that only if the evolution of the three--neutrino system separates into non 
overlapping resonant processes, the transition probabilities are well 
approximated by a wavefunction factorization ansatz.
In general, the use of a phenomenological model is unpractical and arbitrary,
because in this case it is difficult to recognize the dynamics which governs 
the evolution of two (or more) coupled oscillating particles.
Another severe limitation in the application of the bimodal 
factorization consists in the fact that it rests completely inappropriate to 
implement the notion of the rest frame for an unstable 
composite system. This is generally a non--trivially problem because of the
relativistic nature of the theory: a relativistic theory does not allow one to
describe observable states as superpositions of states with different mass and
momentum. Mathematically, this is traced back to the energy dependence of the
matrix Hamiltonian elements. Renouncing the conventional effective Hamiltonian
scheme of Quantum Mechanics, a tantalizing alternative is based on the fermion
mixing transformations  in Quantum Field Theory \cite{QFTmix}. In this picture an
explicit relativistic  description of the essential features of the time
dependent properties of the neutrino system can be introduced by considering
simply the subtleties related to the location of the complex singularities of
the matrix valued propagator. Such propagator's method has the great advantage
to appear natural and indeed, on a first approximation, independent of various
production and decay mechanism. In order to describe the three--neutrino system
in terms of uncoupled propagating neutrinos, we need to diagonalize the matrix
propagator. Generalizing the relevant results obtained for the ${\rm K}^
0-\overline{\rm K}{}^0$ system \cite{CV}, the mixing between on--shell physical
particles in the flavour space is obtained by the dressed inverse matrix
propagator \beq {\Sscr}^{-1}_{\ell h} (p) = 
[\delta_{\ell h} \pslash -\Lambda_{\ell h}] \; ,
\eeq
where $\Lambda_{lh}(p^2)=M_{lh} +
\Sigma_{lh}(p^2)$ consists of the sum of the eventual bare mass--matrix and all
the proper self--energy contributions. The properties of the neutrino system
are characterized by the poles in ${\Sscr}$ which are nearest the physical
sheet. This fermion mixing formalism extends naturally also to massless 
neutrinos, up to subtleties with the massless unitary representations  of the
Poincar\`e group, where only extremal helicities survive as a  consequence of
the privileged role played by the Euclidean little  group.
In the positive energy sector, we have four propagating modes, two 
with positive and two with negative helicity over the chirality 
ratio. The usual four--fold energy degeneracy of spin $1/2$ particles is
removed by a particular form of the self--energy $\Sigma$. For definitiveness,
let us work with a negative helicity  solution and look for a non--trivial
dispersion relation.
The same issues apply to negative-energy
and positive-helicity  solutions.
The fact that the effective ``mass" matrix $\Lambda_{lh}(p^2)=M_{lh} +
\Sigma_{lh}(p^2)$, in general, is momentum dependent does not introduce any
additional complications, in practice \cite{CV}. It is worth noting that the
bare mass matrix $M_{lh}$ is not necessarily diagonal in flavour
space and, in principle, it is neither Hermitian nor symmetric. 
On the other hand, a quantity of fundamental importance in this formalism
is the self--energy matrix. 
The properties of the propagation of the neutrino system are characterized
by the self--energy contributions $\Sigma_{lh}$, from which we obtain  the
generalization of the dispersion relation. The structure of the
neutrino--system self--energy is imposed by the symmetries of the effective
Lagrangian. In absence of a dynamical theory, several specific forms of the
neutrino self--energy can be put forward. In what follows, the neutrino
self--energy is supposed to contribute to the scalar sector. 
Then, some remarks concerning the full propagator are in order. Only the 
resummation of self-energy graphs leads to a Breit-Wigner form which  renders
the Born amplitude finite at the lowest order. Restricting to  vacuum
polarization diagrams, one may obtain the neutrino propagator in the usual way
by summing a geometric series \cite{Inst}. However, there are possible
contributions to the self--energy which arise from  the effects of their mutual
interactions. In principle, we are forced  to include infinite series of higher
order vertex corrections. Thus, the problem becomes non linear and must be
solved in a  self consistent manner by means of the usual approach for the
Schwinger--Dyson equation. In the weak expansion, we expect that other non pole 
singularities in the analytical continuation of the propagator are  small in the
resonant region. Therefore, we follow the point of view  that we are working in
an effective field theory where the various  corrections appear as additional
phenomenological parameters. Indeed, the effective propagator of a neutrino
system incorporates some of the key features required by a gauge theory. As a
matter of fact, field renormalization (and so Green's function renormalization)
should be carried out without altering physical ($S$--matrix) amplitudes.
Usually, the regularizing procedure is obtained by attaching multiplicative
renormalization constants to each free parameter and field, in such a way as
the counterterm Lagrangian as well as the various Green's functions are
automatically gauge invariant. All there constants are fixed by means of
usual on--shell mass renormalization condition with further rules imposed by
Ward identities relating the fermion propagator to the fermion--boson
vertices. Higher order corrections to the vertices, masses and wave
functions, however, could induce considerable complications. The problem consists
in renormalizing a part of a theory where interaction eigenstates are
different from mass eigenstates. The calculability of the Green's functions
is assured by means of a suitable application of these Ward identities.
Therefore, the poles of the renormalized propagator are at the renormalized
eigenmasses. We implement this
condition by requiring
that the physical propagator has the form
\beq\label{propagator}
{\Sscr}_{\alpha \beta} (p)= \delta_{\alpha \beta} \frac {\pslash + \mu_\beta}
{p^2 - \mu^2_\beta}
\eeq
only in the case
the diagonal elements are fixed in such a way that the residues of the
propagators are equal to unity. Of course, the chiral nature of the neutrino
interactions could modify partially this approach. The presence of the masses
couple the right-- and left--components.
The case of massless neutrinos or their
Majorana nature can be included in this discussion with a marginal
modification.  
\vskip 19truemm

\leftline{\bf II. The parametrization of the neutrino mixing.}

\noindent
In the case of weak couplings, the problem to go from mixed to physical states
could be treated just like an algebraic diagonalization procedure with the
determination of the $\Lambda$ eigenvectors \cite{CV}. The matrix $\Lambda$ will
be diagonalized in general by a non--unitary transformation matrix $V$, i.e.
\beq
(V \Lambda V^{-1})_{\beta \alpha}=\mu_\alpha\delta_{\beta
\alpha}. \label{hqd}
\eeq
This matrix $V$ transforms just the initial ``flavour" states
$|\nu_\ell\rangle$ into the ``propagation'' eigenstates
$|\nu_\alpha\rangle$.
Anyway, in the case of scalar self--energy contributions we get 
\beq
{\cal S}_{\alpha \beta} (p) = [\pslash \delta_{\alpha \beta} - N_{\alpha
\beta}]^{-1}= \cases{
{\displaystyle \frac {\delta_{\alpha \beta} D_\beta}{\pslash -\mu_\beta}}
\qquad {\rm with} \qquad D_\beta = [1 - \Sigma^\prime_\beta (s_p) ]^{-1} \cr 
\cr
{\displaystyle \frac {\delta_{\alpha \beta}}{\pslash -s_\beta}} 
\qquad {\rm with} \qquad s_\beta = \mu_\beta + (\pslash -s_p)
\Sigma^\prime_\beta(s_p) \cr}
\eeq
where the dynamical pole of the matrix propagator $s_p$ satisfies the following
relation 
\beq
{\rm det} \; [s_p \delta_{\alpha \beta} - \Lambda_{\alpha \beta}]=0 \; .
\eeq
Here, according to our previous results \cite{CV}, we observe that in general
\beq
N_{\alpha \beta}=\Big[ V \Lambda V^{-1} \Big]_{\alpha \beta} = \delta_{\alpha
\beta} s_\beta = \delta_{\alpha \beta} \Big( m_\beta + \Sigma_\beta \Big)
\eeq 
being
\beq
\delta_{\alpha \beta} \Sigma_\beta = V_{\alpha l} \Sigma_{lh} V^{-1}_{h \beta} \; .
\eeq
The diagonal form of Eq.~(\ref{propagator}) is then recovered only if we
implement on mass shell regularization with the residues equal to unity at the
relevant $s_p$ pole. 

\noindent 
Therefore, the problem consists in finding the
diagonalizing matrix $V$. The only difficulty consists in attaching the
eigenvalue to the right particle. This is performed by comparing the solutions
of the cubic secular equation to the predicted effective mass of each particle.
Once we have identified to which eigenvalue corresponds each state, we can
proceed without any further care about the chosen ordering. It is worth stressing
that the eigenvalues $\mu_{\alpha}$ of $\Lambda$ are the solutions of the secular
equation ${\rm det} \, (\Lambda - \mu I) = 0$ which yields: \beq \mu^3 - ({\rm
tr}\, \Lambda) \mu^2 + {1 \over 2}  \left[({\rm tr}\, \Lambda)^2 - {\rm tr}\,
\Lambda^2 \right] \mu - ({\rm det}\, \Lambda) =0 \; . \eeq
The relevant eigenvalues $\mu_\alpha$ can be assigned to one 
element of the following set
\begin{eqnarray}
\mu_+ & = & \sqrt{u}\left[\sinh{\left(\frac \omega{3}\right)}
+i \sqrt 3 \cosh{\left(\frac \omega{3}\right)}\right] -\frac{b}{3} ,\\ 
\mu_- & = & \sqrt{u}\left[\sinh{\left(\frac \omega{3}\right)}
-i \sqrt 3 \cosh{\left(\frac \omega{3}\right)}\right] -\frac{b}{3},\\      
\mu_0 & = & -2\sqrt{u}\sinh{\left(\frac \omega {3}\right)}-\frac{b}{3}
 ,
\end{eqnarray} 
where
\begin{eqnarray}
 3 u & \equiv & c - \frac{b^2}{3} , \nonumber \\ 
 2 v & \equiv & \frac{2}{27}b^3 - \frac {bc}{3} +d  \qquad {\rm and} \\
 \sinh \omega & \equiv & \frac v{u \sqrt{u}} \nonumber
\end{eqnarray}
being
$b=-\hbox{tr}\Lambda = -\sum_\alpha\mu_\alpha$, 
$c=\frac{1}{2}[(\hbox{tr}\Lambda)^2-\hbox{tr}\Lambda^2] = 
\sum_{\alpha,\beta \atop \alpha\neq \beta} \mu_\alpha 
\mu_\beta$,
and $d=-\hbox{det}\Lambda = - \prod_\alpha \mu_\alpha$.
From these results, we can derive
the transformation matrix $V$ which relates the flavour with
the propagating eigenstates. 
To this end, we can conveniently rewrite the column eigenvectors 
$(a_I,b_I,c_I)^T$
corresponding to each eigenvalue $\mu_I$, $I=\pm, \, 0 $,
in accordance with the following expressions
\begin{eqnarray}
a_I &=& (\mu_I - \Lambda_{22})\Lambda_{13} - \Lambda_{12} \Lambda_{23}
\nonumber \\
b_I &=& -[(\mu_I - \Lambda_{11})\Lambda_{23} + \Lambda_{21} \Lambda_{13}]
\\
c_I &=&  (\mu_I - \Lambda_{11})(\mu_I - \Lambda_{22})+\Lambda_{12}
\Lambda_{21} \nonumber
\end{eqnarray}
where the pedes $I$ labels the subscripts $\pm$, $0$, to assign to the 
eigenstate indices $1,2,3$. 
The last step of the problem consists in the identification of each 
$\mu_\alpha$ to one of the $\mu_I$.
Starting from the elements of $\Lambda$, we can reconstruct the generator $A$ of
$V=e^A$ by means of the relation $e^A = \Lambda_{\rm diag} V \Lambda^{-1}$ where
$\Lambda^{-1}$ can be expressed in terms of the components of the eigenvectors
of $\Lambda$ in the following form:
\begin{equation}
\Lambda^{-1} = 
\frac 1 {{\rm det} \Lambda} \left(
\matrix{
b_- c_0 - b_0 c_- & a_0 c_- - a_- c_0 & a_- b_0 - a_0 b_- \cr
b_0 c_+ - b_+ c_0 & a_+ c_0 - a_0 c_+ & a_0 b_+ - a_+ b_0 \cr
b_+ c_- - b_- c_+ & a_- c_+ - a_+ c_- & a_+ b_- - a_- b_+ \cr}
\right) \;.
\end{equation}
In particular, assuming that $V$ is unitary ($A^{\dag}=-A$) and using the
Cayley--Hamilton decomposition \cite{VEA}, we can write
\begin{equation}
V=e^A=I + \left( \frac{\sin \Xscr}{\Xscr}\right)A + 
\frac 1 2 \left[\frac{\sin ({\Xscr}/2)}{{\Xscr}/2}\right]^2A^2
\end{equation}
in terms of the elements of the matrix $A$. If $A$ is real, neglecting any
CP--effects, it must be of the form
\begin{equation}
A = \left(
\matrix{
0 & \chi_1 & \chi_3 \cr
-\chi_1 & 0 & -\chi_2 \cr
-\chi_3 & \chi_2 & 0 \cr}
\right) 
\end{equation}
and consequently, $V$ will be given by
\begin{equation}
V = \left(
\matrix{
1- {\displaystyle\frac{\Xscr^2-\chi_2^2} 2} & 
\chi_1+{\displaystyle\frac 1 2}  \chi_2\chi_3  &
\chi_3-{\displaystyle\frac 1 2}  \chi_1\chi_2 \cr
\cr 
-\chi_1+{\displaystyle\frac 1 2}  \chi_2\chi_3 &
1-{\displaystyle\frac{\Xscr^2-\chi_3^2} 2} & 
-\chi_2-{\displaystyle\frac 1 2}  \chi_1\chi_3 \cr
\cr 
-\chi_3-{\displaystyle\frac 1 2}  \chi_1\chi_2 &
\chi_2-{\displaystyle\frac 1 2}  \chi_1\chi_3 & 
1-{\displaystyle\frac{\Xscr^2-\chi_1^2} 2} \cr} 
\right) \; ,
\end{equation}
where $\Xscr=\sqrt{\chi^2_1 + \chi^2_2 + \chi^2_3}$.
To the end of disentangling electron 
neutrino $\nu_e$ sector, we can get an approximated parametrization of $V$ in
terms of only the real parameter $\chi$ 
\begin{equation}
\vert V \vert \simeq \left(
\matrix{
1 & \chi^3  & \chi^2 \cr  
\chi^3 & 1 & \chi \cr
\chi^2 & \chi & 1 \cr} 
\right) \; ,
\end{equation}
if we consider $\chi_1 \simeq \chi^3, \, 
\chi_2 \simeq \chi, \, \chi_3 \simeq \chi^2$
consistently with a large angle oscillations of $\nu_\mu$ 
predominantly into $\nu_\tau$. 
Supposing an analogy among quarks and leptons as a basic input, we expect $\chi$
of the same order of magnitude of the Cabibbo angle $\lambda$. Indeed, a value 
$\chi \simeq 1.5 \lambda$ seems to provide a good agreement with some recent  
best fits \cite{Krastev}.  
This result can be best appreciated by considering the parametrization of the
mixing matrix of the quark sector
\begin{equation}
V_{CKM} \simeq
\left(
\matrix{
1 & \lambda  & \lambda^3 \cr  
\lambda & 1 & \lambda^2 \cr
\lambda^3 & \lambda^2 & 1 \cr} 
\right) \; ,
\end{equation}
in terms of the Cabibbo angle $\lambda \simeq 0.22$, just omitting other unitarization 
parameters \cite{Pavia}. The observed quark mass hierarchies
\begin{equation}
\frac {m_u}{m_t} \sim \lambda^8 \; ,\qquad \frac {m_c}{m_t} \sim \lambda^4 
\; , \qquad 
\frac {m_d}{m_b} \sim \lambda^4 \; ,\qquad \frac {m_s}{m_b} \sim \lambda^2 
\; ,
\end{equation}
suggest to interpret this pattern as due to some family regularities
in the Yukawa sector and fundamentally to a mass giving mechanism. 
The observed quark mixing
parameters are given at the scale of the neutral gauge boson $Z^0$. 
On the other hand, the rescaling effects
do not seem to affect substantially these mixing features \cite{denner}. 
Also in the lepton sector, the renormalization effects, which control the 
redefinition 
among weak and ``strong" eigenstates, are expected negligible
in many extension of the Standard Model \cite{kniehl}.
The texture structure for quark masses induces to
ask whether a similar texture could be
used for neutrino masses. In fact, 
a similar inspection in the pattern of the charged lepton ratios
\begin{equation}
\frac {m_e}{m_\tau} \sim \chi^8 \qquad \frac {m_\mu}{m_\tau} \sim \chi^4
\end{equation} provides
an interesting prediction for neutrino mass ratios
\begin{equation}
\frac {m_{\nu_e}}{m_{\nu_\tau}} \sim \chi^4 \qquad \frac {m_{\nu_\mu}}
{m_{\nu_\tau}} \sim \chi^2 .
\end{equation}  
Indeed, the explanation of this common hierarchical texture and the origin
of the Cabibbo suppressions of the Yukawa couplings is now poorly understood, and
it remains one of the most outstanding problems of particle physics. 
Today, many different mass matrix morphologies appear consistent with our
approach. Usually, the full
neutrino mixing matrix $V$ is addressed analogously to the 
Cabibbo-Kobayashi-Maskawa quark mixing matrix.
However, we can recognize that the lepton 
sector is largely different from the quark sector. Apparently, this 
difference seems to reside in the 
fact that the $\nu_\mu$--$\nu_\tau$ mixing is very large compared to 
$V_{cb}^{CKM}$. Indeed, the main issue consists in the fact that most of
the neutrino masses are extremely small, if not null, compared with the 
charged lepton masses. Furthermore, we must not forget a fundamental
difference between the  mixing of neutrinos and quarks.  
Only if the lepton
numbers $L_e$,  $L_\mu$ and $L_\tau$ are conserved, massive neutrinos can be
treated  on the same foot of the quarks. In this case a massive neutrino 
reveals its Dirac nature and differs from an antineutrino by an  opposite value
of the lepton charge. Whatever the lepton numbers are  not conserved, neutrino
oscillations may be considered only in the  scheme of mixing of two--component
Majorana neutral particles \cite{kniehl}.
This ambiguity on the Dirac or Majorana nature of the neutrino 
masses enlarges the variety of the possible parametrizations of the
mixing matrix $V$.
Without further specifications,
we have not found convenient to describe the neutrino mixing
with a modification of earlier representations
advocated in the case of the Euler-type quark mixing~\cite{Pavia}. 
Patterns of neutrino masses and mixing as suggested by the present
experimental evidences seem to support a morphology quite different 
from the Cabibbo predominance emerging in the quark sector.
Independently of all other features,
the physical significance of $V$ imposes that its elements must be 
invariant under their rephasing transformations. 
Unfortunately,
this invariance makes each matrix element of $V$ physically
meaningless. 
Of course, some phases can be eliminated by means of a redefinition
of the neutrinos. But, 
physical observables cannot depend on field redefinitions. In the
mass--eigenstate basis, the only field redefinitions are 
unitary transformations which leave unchanged the diagonal mass matrix. 
Thus, the corresponding observables can only depend on quantities 
invariant under these transformations. The simplest of these 
quantities are just the moduli of the $V$--matrix elements. 
Besides moduli, only a recombination of matrix elements of the kind of 
those proposed by Jarlskog can be identified as an invariant~\cite{Martin}.
Therefore, although this is not
the common  practice, a convenient parametrization of the mixing matrix consists
of a minimum set of four independent invariants to select among the squared
moduli and the Jarlskog--like plaquette phases. The choice of the set of basic
invariants for three generations should be dictated by present experimental
data. Our modest proposal relies on the disentangling of the electron 
neutrino $\nu_e$ sector. 
According to that, the most feasible invariants which
we can choose are
$J={\rm Im} (V_{21}V_{32}V_{22}^*V_{31}^*)$ and three other basic moduli
among $\vert V_{21}\vert^2, \vert V_{32}\vert^2, \vert V_{31}\vert^2$ or 
$\vert V_{22}\vert^2$. This choice is easily checked to generate a reliable
three--generation mixing matrix. There are of course many other possible
choices of basic invariants which should be dictated by the availability and
precision of next experimental data. Nevertheless, any other proposal should
take into account that the most accessible information that we have about $V$
can be extracted by the moduli. In fact, observable quantities can be extracted
from the transition probabilities which depend on moduli. 
The amplitude of Eq.~(\ref{amplitude}) for the transition of one neutrino flavour
into another can be rewritten as
\beq
{\cal A}_{lh} = \sum_{\alpha, \beta} V_{\alpha h} \Delta^{-1}_{\alpha \beta}
V^*_{\beta l} 
\eeq
by means of the inverse matrix $\Delta^{-1}_{\alpha \beta}$ for
the propagating eigenstates,
where again $l,h$ denote the flavour-- and $\alpha , \beta$ the
propagating--eigenstates and $V_{\alpha l}$ are the elements of the neutrino
mixing matrix emerging in the redefinition of Eq.~(\ref{transf}).
The probability of the flavour transitions as a function of space--time are then
given by
\beq
{\cal P}(\nu_l \rightarrow \nu_h) = \sum_{\alpha} \Big \vert V_{\alpha h} 
\Delta^{-1}_{\alpha \alpha} V^*_{\alpha l} \Big \vert^2 + \sum_{\alpha \neq
\beta} V_{\alpha h} V^*_{\alpha l} \Delta^{-1 {\ast}}_{\alpha \alpha} 
\Delta^{-1}_{\beta \beta}V^*_{\beta h}V_{\beta l}
\eeq
and, using explicitly the proposed invariants for the
parametrization of the mixing matrix,
the time--dependent probability for the $\nu_e \rightarrow \nu_\mu$ conversion
can be written as
\begin{eqnarray}
{\cal P}(\nu_e \rightarrow \nu_\mu) &=& \sum_{\alpha} \vert V_{\alpha
\mu}\vert^2  \vert V_{\alpha e}\vert^2\vert\Delta^{-1}_{\alpha \alpha}\vert^2 
-2{\rm Re}\sum_{\alpha \leq \beta} \Big[V_{\alpha e} V_{\beta \mu} V^*_{\alpha
\mu}V^\ast_{\beta e}\Delta^{-1}_{\alpha \alpha}\Big(\Delta^{-1}_{\beta
\beta}\Big)^\ast\Big] \sim \nonumber \\
&\sim & |{\cal W}_{11}|^2 \chi^6 + |{\cal W}_{22}|^2 \chi^6 + 
|{\cal W}_{33}|^2 \chi^6 + \nonumber \\
&+& 2\; {\rm Re} \Big[{\cal W}_{11} {\cal W}_{22}^* \chi^6 +
{\cal W}_{11}{\cal W}_{33}^* \chi^6 + {\cal W}_{22}{\cal W}_{33}^* \chi^6 \Big]
\sim  \chi^6 \; ,  
\end{eqnarray}
where the quantities ${\cal W}_{ii}$ appear in Eq.~(\ref{prop}), and are of the order of
unity in absence of regeneration effects \cite{CV}.
This estimate seems fairly consistent with
the recent results of the LSND and CHOOZ collaborations
\cite{LSND}.
Analogously, it is possible to obtain an expression for the
probability for the $\nu_\mu \leftrightarrow \nu_\tau$ 
transition
\begin{eqnarray}
{\cal P}(\nu_\mu \rightarrow \nu_\tau) &\sim & |{\cal W}_{11}|^2 \chi^{10} + 
|{\cal W}_{22}|^2 \chi^2 + |{\cal W}_{33}|^2 \chi^2 + \nonumber \\
&+& 2\; {\rm Re} \Big[{\cal W}_{11}{\cal W}_{22}^* \chi^6 +
{\cal W}_{11}{\cal W}_{33}^* \chi^7 +{\cal W}_{22}{\cal W}_{33}^* \chi^2 \Big]
\sim \chi^2 \end{eqnarray}
where the last approximation refers to the real parameter $\chi$, we 
introduced previously.
On the other hand, in general the relevant time--integrated probability is given 
by
\beq {\widehat {\cal P}}(\nu_l \rightarrow \nu_h) = \sum_{\alpha} \Big \vert 
V_{\alpha h}  A_{\alpha \alpha} V^*_{\beta l} \Big \vert^2 + 
\sum_{\alpha \neq \beta} V_{\alpha h} V^*_{\alpha l} B_{\alpha \beta} 
V^*_{\beta h}V_{\beta l} \; . 
\eeq
In a wavepacket analysis of the oscillation problem, the stationary phase method
can provide us the necessary technicality to evaluate the relevant quantities
\begin{eqnarray}
A_{\alpha \alpha} &=& {\displaystyle \int_0^\infty} \vert \Delta_{\alpha
\alpha}^{-1}\vert^2 dt   
  \nonumber \\ 
&& \\
B_{\alpha \beta} &=& {\displaystyle \int_0^\infty }\Delta_{\alpha
\alpha}^{-1\ast} \Delta_{\beta \beta}^{-1} dt
\nonumber \end{eqnarray}
which are obtained assuming the 
independent resonant propagation of the three neutrinos.
Furthermore, neglecting any coherent effect, the probability of the flavour
conversion as a function of distance can still be written as 
\beq
{\widehat {\cal P}}(\nu_l \rightarrow \nu_h) = \delta_{\alpha \beta} -4
\sum_{\alpha > \beta} V_{\alpha h} V^*_{\alpha l} 
V^*_{\beta h}V_{\beta l} \sin^2(\varphi_M^{\alpha \beta} + 
\varphi_\Sigma^{\alpha \beta}) \; . 
\eeq
where $\varphi_M^{\alpha \beta} = \frac{\pi L}{\delta^{\alpha \beta}_M}$ 
has a characteristic 
length $\delta_M^{\alpha \beta} \simeq \frac{4\pi E}{\Delta m_{\alpha\beta}}$ which is
induced by a mass difference $\Delta m_{\alpha\beta} = m_\alpha - 
m_\beta$ and is proportional to the energy, at least at the first 
approximation \cite{cohe}. The interacting phase $\varphi_\Sigma^{\alpha \beta} 
=\frac{2\pi}{\vert \delta^{\alpha \beta}_\Sigma \vert}$ will appear, for instance,
whenever there 
exists a non vanishing mixing angle induced by flavour dependent 
dispersion relations.
The mixing parameters can then be expressed in terms of these oscillating 
probabilities.

\vskip 10truemm
\leftline{\bf III. Conclusions}

\noindent
In the vacuum, in absence of external interactions,
the lepton flavour mixing is usually read off by means of the nontrivial 
neutrino mass matrix $M_{\ell h}$, in the basis of the diagonal charged 
lepton mass matrix.
The present available information does not seem to select 
the texture of the neutrino mixing matrix \cite{Fogli}.
Only assuming a democratic bimaximal mixing we are lead to a defined  
structure~\cite{Feruglio}, which can describe both solar and 
atmospheric neutrino oscillations. This matrix structure implies that the 
electron neutrino $\nu_e$ does not oscillate at the atmospheric scale,
in contrast to the large mixing $\nu_\mu \leftrightarrow \nu_\tau$
emerging from the results of the Super Kamiokande Collaboration~\cite{SKam}.
\noindent
In any case, the family symmetries 
which constrain this behaviour,
will probably lie outside of
the realm of the Standard Model and can be gauged in order to  generate neutrino
mass-splittings. Typically, this attempt gives rise to large custodial violating
effects \cite{Lusi} \ 
which do not protect from flavour changing neutral currents, and
induce a violation of the Standard Model GIM mechanism \cite{ZFCNC}. So that we
are faced out with the presence of corresponding pseudo Goldstone bosons, called Majoron or
familons, when associated with broken lepton numbers. 
Once these pseudoscalar bosons are
generated, they can change in turn the relative propagating features of the different
neutrino species. As a result, neutrinos may exhibit oscillation patterns for the possible effects
caused by these Majoron fields~\cite{Bento}. 

\noindent
However, the problem to arrange a sensible pattern of neutrino masses and 
mixings is connected with the breaking of the $B-L$ invariance. 
This problem has undoubted connection with
the non abelian nature of chiral symmetry breaking in Quantum 
Chromodynamics.
In our case, the independent
conservation of the  lepton numbers $L_e$, $L_\mu$ and  $L_\tau$, must be
associated to global symmetries, which cannot be gauged in the Standard Model,
but which are known to be unstable in presence of a suitable background. For
example, it is well known that such global symmetries are unstable against 
gravitational effects \cite{QGrav} with potentially disastrous consequences. 
The manifestation of the breaking consists in an effective interaction which 
breaks the symmetry in the
low energy sector of the theory through a lepton-number violating coupling.
Then, there is some additional softly broken local symmetry associated 
with torsion. In presence of a curved space--time with torsion, 
the divergence of the  leptonic currents gets
new contributions in addition to those coming  from the electroweak fields
\cite{Torsion}. Therefore, the issues of origin and stability of lepton numbers
are  rooted in the realm of a convenient unification of gauge and  gravitational
interactions at high scales.  A common feature of these  theories is the
presence of compactified extra dimensions. It is not a  strange idea that the
presence of extra dimensions can play a role  for neutrino mass generation
\cite{DDIM}. In this kind of theories with a string ground state manifold, the
breaking of global symmetries is associated with some moduli fields  which
cannot avoid a mass at higher energy scales. String moduli have  only
nonrenormalizable couplings to ordinary physical fields with the typical  range
close to the Planck scale.  Among the many kind of moduli fields, the  dilaton
is particularly important because it determines the gauge  couplings. The
dilaton arises in theories with spontaneously broken  scale invariance and
obtains its mass from the effects due to the  conformal anomaly.  Such effects
at low energies induce very peculiar couplings. While the dilaton interacts with
ordinary matter  universally, moduli fields can interact with different
interaction  strength parameters \cite{Moduli}. 
This scenario is particularly appealing
because it involves  gravitational effects which can be experimentally tested in
the working  colliders and those that will turn on during the coming years. 
The formalism we have developed will then become an important tool to study 
these complications.

\vfill\eject


\begin{thebibliography}{100}

\bibitem{BP} B. Pontecorvo, Sov. Phys. JETP {\bf 26}, 984 (1968).
For an update review see:
S. M. Bilenky, C. Giunti and W. Grimus, e-preprint archive hep-ph/9812360.

\bibitem{CV} D. Cocolicchio, Nuovo Cimento, {\bf A111}, 1197 (1998);
D. Cocolicchio and M. Viggiano,
``{\it The Quantum Field Theory of the Kaon Oscillations}'',
preprint Sezione INFN Milano IFUM 600-FT/97 (e-preprint archive
hep-ph/9712480), to appear in Nuovo Cimento,  {\bf A112} (1999).

\bibitem{SKam} Y. Fukuda {\it et al.}, Super Kamiokande Collab.,
Phys. Rev. Lett. {\bf 81}, 1562 (1998).

\bibitem{LSND} C. Athanassopoulos {\it et al.}, LSND Collab., Phys. 
Rev. Lett. {\bf 81}, 1774 (1998). 

\bibitem{SSM} J. N. Bahcall, Phys. Lett. {\bf B433}, 1 (1998).

\bibitem{CHOOZ} M. Apollonio {\it et al.}, CHOOZ Collab., Phys. Lett. {\bf B420},
397 (1998).

\bibitem{MSW} L. Wolfenstein, Phys. Rev. {\bf D17}, 2369 (1978); S. P. 
Mikheyev and A. Yu. Smirnov, Sov. J. Nucl. Phys. {\bf 42}, 913 (1985).

\bibitem{Magnus} W. Magnus, Commun. Pure Appl. Math. {\bf 7}, 649 (1954).

\bibitem{wilcox} R. M. Wilcox, J. Math. Phys. {\bf 8}, 962 (1967).

\bibitem{BCH} M. Lutzky, J. Math. Phys. {\bf 9}, 1125 (1968); J. A. 
Oteo, J. Math. Phys. {\bf 32}, 414 (1991).

\bibitem{birula} D. Cocolicchio and M. Viggiano, Int. J. Theor.
Phys. {\bf 37}, 2079 (1998).

\bibitem{baldini} A. Baldini and G.F. Giudice, Phys. Lett. {\bf B186}, 211 (1987);
J. C. D'Olivo and J. A. Oteo, Phys. Rev. {\bf D54}, 1187 (1996).

\bibitem{cohe} B. Kaiser, Phys. Rev. {\bf D24}, 110 (1981);
K. Kiers and N. Weiss, Phys. Rev. {\bf D57}, 3091 (1998);
W. Grimus, P. Stockinger and S. Mohanty, 
Phys. Rev. {\bf D59}, 013011 (1999);
A. Ioannisian and A. Pilaftsis, Phys. Rev. {\bf D59}, 053003 (1999) .

\bibitem{dens} R. A. Harris and L. Stodolsky, Phys. Lett. {\bf B116}, 464
(1982); C. P. Burgess and D. Michaud, Ann. Phys. (NY) {\bf 256}, 1 (1997);
M. Sirera and A. Perez, Phys. Rev. {\bf D59}, 125011 (1999).

\bibitem{QFTmix} J. F. Donoghue, Phys. Rev. {\bf D19}, 2772 (1979);
M. Capdequi Peyran\`ere and M. Talon, Nuovo Cimento 
{\bf 75A}, 205 (1983);
M. Blasone and G. Vitiello,  Ann. Phys. (N.Y.) {\bf 244}, 283 (1995); 
E {\it ibidem} {\bf 249}, 363 (1996);
E.Alfinito, M.Blasone, A.Iorio and G.Vitiello, 
Phys. Lett. {\bf B 362}, 91 (1995); K. Fujii, C. Habe and T. 
Yabuki, Phys. Rev. {\bf D59}, 113003 (1999).

\bibitem{Inst} D. Cocolicchio, Phys. Rev. {\bf D57}, 7251 (1998).

\bibitem{VEA} G. Dattoli, C. Mari and A. Torre, 
Nuovo Cimento {\bf 108B}, 61 (1993).

\bibitem{Krastev} J. N. Bahcall, P. I. Krastev and A. Y. Smirnov, Phys. Rev.
{\bf D58}, 096016 (1998); T. Fukuyama, K. Matsuda and H. Nishiura, Mod. Phys. Lett.
{\bf A13}, 2279 (1998);
G. Barenboim and F. Scheck, Phys. Lett. {\bf B450}, 189 (1998);
T. Teshima and T. Sakai, Prog. Theor. Phys. 
{\bf 101}, 147 (1999); G. Fogli, E. Lisi, A. Marrone and G. Scioscia, 
Phys. Rev. {\bf D59}, 033001 (1999).

\bibitem{Pavia} D. Cocolicchio,
``{\it $CP$--asymmetries in $B$ decays}'', Proc.
{\it Advanced Study Conference on Heavy Flavours},
ed. G. Bellini et al.\ (Ed. Frontieres, 1993), p. 367;
H. Fritzsch and Z.-z. Xing, Phys. Rev. {\bf D57}, 594 (1998).

\bibitem{denner} A. Denner and T. Sack, Nucl. Phys. {\bf B347}, 203 (1990);
H. Fusaoka and Y. Koide, Phys. Rev. {\bf D57}, 3986 (1998); 
C. Balzereit, Th. Hansmann, T. Mannel and B. Pl\"umper, 
preprint Univ. Karlsruhe
TTP-98-34, e-preprint archive hep-ph/9810350.

\bibitem{kniehl} B. A. Kniehl and A. Pilaftsis, Nucl. Phys. {\bf B474}, 286 (1996).

\bibitem{Martin} G. C. Branco and L. Lavoura, Phys. Lett. {\bf B208}, 123 
(1988); G. Auberson, A. Martin and G. Mennessier,
Commun. Math. Phys. {\bf 140}, 523 (1991).

\bibitem{Fogli} G. Fogli, E. Lisi and D. Montanino, Astropart. Phys. 
{\bf 9}, 119 (1998).

\bibitem{Feruglio}  V. Barger, S. Pakvasa, T.~J. Weiler and K. 
Whisnant, Phys. Lett. {\bf B437}, 107 (1998);
G. Altarelli and F. Feruglio, Phys. Lett. {\bf B439}, 112 (1998).

\bibitem{Lusi} 
D. Grasso, M. Lusignoli and M. Roncadelli, Phys. Lett. {\bf B288}, 140 (1992); 
R. Barbieri, L. J. Hall, G. L. Kane, G. G. Ross, e-preprint archive
hep-ph/9901228.

\bibitem{ZFCNC} J. Bernabeu {\it et al.}, 
Phys. Rev. Lett. {\bf 71}, 2695 (1993), Phys. Lett. {\bf B351}, 235 (1995); 
P. Langacker and D. London, Phys. Rev. {\bf D38}, 886 (1988).

\bibitem{Bento} L. Bento, Phys. Rev. {\bf D57}, 583 (1998);
Phys. Rev. {\bf D59}, 015013 (1999). 

\bibitem{QGrav} S. Giddings and A. Strominger, Nucl. Phys. {\bf B306}, 890
(1988); S. Coleman and K. Lee, Nucl. Phys. {\bf B325}, 387 (1989); 
L. F. Abbot and M. B. Wise, Nucl. Phys. {\bf B329}, 387 (1990);
R. Kallosh and A. Linde, Phys. Rev. {\bf D52}, 912 (1995).

\bibitem{Torsion} A. Dobado and A. L. Maroto,
e-preprint archive hep-ph/9705434.

\bibitem{DDIM} N.Arkani-Hamed, S. Dimopoulos, G. Dvali and J. M. 
Russel, e-preprint archive hep-ph/9811448; K. R. Dienes, E. Dudas and T. Gherghetta, 
e-preprint archive hep-ph/9811428.

\bibitem{Moduli} K. Kobayakawa, Y. Sato and S. Tanaka, Phys. Rev. {\bf 
D54}, 1204 (1996).

\end{thebibliography}
\end{document}